\documentclass[trackchanges]{aastex701}
\usepackage{amsmath}
\usepackage{ulem}
\usepackage{subfigure}
\usepackage{subfiles}
\usepackage{booktabs}
\usepackage{color}
\usepackage{graphicx}
\usepackage[export]{adjustbox}
\usepackage{CJKutf8}
\begin{document}

\title{A Robust Geometric Distortion Solution for Main Survey Camera of CSST}
\correspondingauthor{Jundan Nie, Chao Liu}
\email{jdnie@nao.cas.cn, chaoliu@bao.ac.cn}

\author[0009-0001-0987-9942]{Yibo Yan\
\begin{CJK}{UTF8}{gbsn}(闫一波)\end{CJK}}
\affiliation{National Astronomical Observatories, Chinese Academy of Sciences, Beijing, 100101, People's Republic of China}
\affiliation{University of Chinese Academy of Sciences, Beijing, 100049, People's Republic of China}
\email{ybayan@bao.ac.cn}

\author[0000-0002-3616-9268]{You Wu\
\begin{CJK}{UTF8}{gbsn}(吴优)\end{CJK}}
\affiliation{National Astronomical Observatories, Chinese Academy of Sciences, Beijing, 100101, People's Republic of China}
\email{wuyou@nao.cas.cn}

\author[0000-0001-6590-8122]{Jundan Nie\
\begin{CJK}{UTF8}{gbsn}(聂俊丹)\end{CJK}}
\affiliation{National Astronomical Observatories, Chinese Academy of Sciences, Beijing, 100101, People's Republic of China}
\affiliation{University of Chinese Academy of Sciences, Beijing, 100049, People's Republic of China}
\email{jdnie@nao.cas.cn}

\author[0000-0002-8531-5161]{Tianmeng Zhang\
\begin{CJK}{UTF8}{gbsn}(张天萌)\end{CJK}}
\affiliation{National Astronomical Observatories, Chinese Academy of Sciences, Beijing, 100101, People's Republic of China}
\affiliation{University of Chinese Academy of Sciences, Beijing, 100049, People's Republic of China}
\email{zhangtm@nao.cas.cn}

\author[0000-0002-1802-6917]{Chao Liu\
\begin{CJK}{UTF8}{gbsn}(刘超)\end{CJK}}
\affiliation{National Astronomical Observatories, Chinese Academy of Sciences, Beijing, 100101, People's Republic of China}
\affiliation{University of Chinese Academy of Sciences, Beijing, 100049, People's Republic of China}
\affiliation{Institute for Frontiers in Astronomy and Astrophysics of Beijing Normal University, Beijing, 100875, People's Republic of China}
\affiliation{Zhejiang Lab, Hangzhou, 311121, People's Republic of China}
\email{chaoliu@bao.ac.cn}

\author{Zhang Ban\begin{CJK}{UTF8}{gbsn}(班章)\end{CJK}}
\affiliation{Changchun Institute of Optics, Fine Mechanics and Physics, Chinese Academy of Sciences, Changchun, 130033, People's Republic of China}
\email{banzhang0711@163.com}

\author[0000-0002-9494-0946]{Zihuang Cao\begin{CJK}{UTF8}{gbsn}(曹子皇)\end{CJK}}
\affiliation{National Astronomical Observatories, Chinese Academy of Sciences, Beijing, 100101, People's Republic of China}
\email{zhcao@nao.cas.cn}

\author[0000-0003-4546-8216]{Wei Du\begin{CJK}{UTF8}{gbsn}(杜薇)\end{CJK}}
\affiliation{National Astronomical Observatories, Chinese Academy of Sciences, Beijing, 100101, People's Republic of China}
\email{wdu@nao.cas.cn}

\author{Yuedong Fang\begin{CJK}{UTF8}{gbsn}(方越东)\end{CJK}}
\affiliation{University Observatory, Faculty of Physics, Ludwig-Maximilians-Universität, Scheinerstr. 1, 81679 Munich, Germany}
\email{Yuedong.Fang@physik.lmu.de}

\author[0000-0003-3317-4771]{Yi Hu\begin{CJK}{UTF8}{gbsn}(胡义)\end{CJK}}
\affiliation{National Astronomical Observatories, Chinese Academy of Sciences, Beijing, 100101, People's Republic of China}
\email{huyi@nao.cas.cn}

\author[0000-0003-4211-851X]{Guoliang Li\begin{CJK}{UTF8}{gbsn}(李国亮)\end{CJK}}
\affiliation{Purple Mountain Observatory, Chinese Academy of Sciences, Nanjing, Jiangsu, 210023, People's Republic of China}
\email{guoliang@pmo.ac.cn}

\author{Xiaobo Li\begin{CJK}{UTF8}{gbsn}(李晓波)\end{CJK}}
\affiliation{Changchun Institute of Optics, Fine Mechanics and Physics, Chinese Academy of Sciences, Changchun, 130033, People's Republic of China}
\email{lixiaobo104@163.com}

\author[0000-0003-4952-3008]{Chenxiaoji Ling\begin{CJK}{UTF8}{gbsn}(凌晨晓骥)\end{CJK}}
\affiliation{National Astronomical Observatories, Chinese Academy of Sciences, Beijing, 100101, People's Republic of China}
\email{lcxj@bao.ac.cn}

\author[0000-0002-3134-9526]{Jiaqi Lin\begin{CJK}{UTF8}{gbsn}(林家琪)\end{CJK}}
\affiliation{Shanghai Astronomical Observatory, Chinese Academy of Sciences, 80 Nandan Road, Shanghai, 200030, People's Republic of China}
\affiliation{School of Physics and Astronomy, Sun Yat-sen University, Zhuhai, 519082, People's Republic of China}
\affiliation{CSST Science Center for the Guangdong-Hong Kong-Macau Greater Bay Area, Zhuhai, 519082, People's Republic of China}
\email{linjq63@mail2.sysu.edu.cn}

\author[0000-0002-0409-5719]{Dezi Liu
\begin{CJK}{UTF8}{gbsn}(刘德子)\end{CJK}}
\affiliation{South-Western Institute for Astronomy Research, Yunnan University, Kunming, 650500, China}
\email{adzliu@ynu.edu.cn}

\author{Yu Luo
\begin{CJK}{UTF8}{gbsn}(罗煜)\end{CJK}}
\affiliation{School of Physics and Electronics, Hunan Normal University, 36 Lushan Road, Changsha,
410081, People's Republic of China}
\email{luoyupmo@gmail.com}

\author[0000-0002-6077-6287]{Bin Ma\begin{CJK}{UTF8}{gbsn}(马斌)\end{CJK}}
\affiliation{School of Physics and Astronomy, Sun Yat-sen University, Zhuhai, 519082, People's Republic of China}
\affiliation{CSST Science Center for the Guangdong-Hong Kong-Macau Greater Bay Area, Zhuhai, 519082, People's Republic of China}
\email{mabin3@mail.sysu.edu.cn}

\author{Xianmin Meng\begin{CJK}{UTF8}{gbsn}(孟宪民)\end{CJK}}
\affiliation{National Astronomical Observatories, Chinese Academy of Sciences, Beijing, 100101, People's Republic of China}
\email{mengxm@nao.cas.cn}

\author[0000-0003-3243-464X]{Juanjuan Ren\begin{CJK}{UTF8}{gbsn}(任娟娟)\end{CJK}}
\affiliation{National Astronomical Observatories, Chinese Academy of Sciences, Beijing, 100101, People's Republic of China}
\email{jjren@nao.cas.cn}

\author[0000-0003-2015-777X]{Li Shao\begin{CJK}{UTF8}{gbsn}(邵立)\end{CJK}}
\affiliation{National Astronomical Observatories, Chinese Academy of Sciences, Beijing, 100101, People's Republic of China}
\email{shaoli@nao.cas.cn}

\author[0000-0003-3347-7596]{Hao Tian
\begin{CJK}{UTF8}{gbsn}(田浩)\end{CJK}}
\affiliation{National Astronomical Observatories, Chinese Academy of Sciences, Beijing, 100101, People's Republic of China}
\email{tianhao@nao.cas.cn}

\author[0000-0001-5912-7522]{Chengliang Wei \begin{CJK}{UTF8}{gbsn}(韦成亮)\end{CJK}}
\affiliation{Purple Mountain Observatory, Chinese Academy of Sciences, Nanjing, Jiangsu, 210023, People's Republic of China}
\email{chengliangwei@pmo.ac.cn}

\author[0000-0003-2477-6092]{Peng Wei\begin{CJK}{UTF8}{gbsn}(魏鹏)\end{CJK}}
\affiliation{National Astronomical Observatories, Chinese Academy of Sciences, Beijing, 100101, People's Republic of China}
\email{weipeng01@bao.ac.cn}

\author[0009-0008-7783-945X]{Shoulin Wei\begin{CJK}{UTF8}{gbsn}(卫守林)\end{CJK}}
\affiliation{Faculty of Information Engineering and Automation, Kunming University of Science and Technology, Kunming, 650500, People's Republic of China}
\email{weishoulin@kust.edu.cn}

\author[0009-0004-2243-8289]{Yun-Ao Xiao\begin{CJK}{UTF8}{gbsn}(肖云奥)\end{CJK}}
\affiliation{National Astronomical Observatories, Chinese Academy of Sciences, Beijing, 100101, People's Republic of China}
\affiliation{University of Chinese Academy of Sciences, Beijing, 100049, People's Republic of China}
\email{xiaoya@nao.cas.cn}

\author{Zhou Xie\begin{CJK}{UTF8}{gbsn}(谢洲)\end{CJK}}
\affiliation{Center for Astrophysics and Great Bay Center of National Astronomical Data Center, Guangzhou University, Guangzhou, Guangdong, 510006, People's Republic of China}
\email{jnk\_xz@outlook.com}

\author[0000-0002-9728-1552]{Su Yao\begin{CJK}{UTF8}{gbsn}(姚苏)\end{CJK}}
\affiliation{National Astronomical Observatories, Chinese Academy of Sciences, Beijing, 100101, People's Republic of China}
\email{yaosu@bao.ac.cn}

\author[0000-0002-7210-8104]{Yan Yu\begin{CJK}{UTF8}{gbsn}(吁彦)\end{CJK}}
\affiliation{National Astronomical Observatories, Chinese Academy of Sciences, Beijing, 100101, People's Republic of China}
\affiliation{School of Physics and Astronomy, Sun Yat-sen University, Zhuhai, 519082, People's Republic of China}
\email{yuyan35@mail2.sysu.edu.cn}

\author[0009-0009-9256-9248]{Shengwen Zhang\begin{CJK}{UTF8}{gbsn}(张圣文)\end{CJK}}
\affiliation{National Astronomical Observatories, Chinese Academy of Sciences, Beijing, 100101, People's Republic of China}
\affiliation{University of Chinese Academy of Sciences, Beijing, 100049, People's Republic of China}
\email{swzhang@bao.ac.cn}

\author[0000-0001-7314-4169]{Xin Zhang \begin{CJK}{UTF8}{gbsn}(张鑫)\end{CJK}}
\affiliation{National Astronomical Observatories, Chinese Academy of Sciences, Beijing, 100101, People's Republic of China}
\email{zhangx@nao.cas.cn}

\author[0009-0002-6783-7510]{Bowei Zhao\begin{CJK}{UTF8}{gbsn}(赵博伟)\end{CJK}}
\affiliation{National Astronomical Observatories, Chinese Academy of Sciences, Beijing, 100101, People's Republic of China}
\affiliation{University of Chinese Academy of Sciences, Beijing, 100049, People's Republic of China}
\email{zhaobw@nao.cas.cn}

\author[0000-0002-4135-0977]{Zhimin Zhou\begin{CJK}{UTF8}{gbsn}(周志民)\end{CJK}}
\affiliation{National Astronomical Observatories, Chinese Academy of Sciences, Beijing, 100101, People's Republic of China}
\email{zmzhou@bao.ac.cn}

\author[0000-0002-6684-3997]{Hu Zou\begin{CJK}{UTF8}{gbsn}(邹虎)\end{CJK}}
\affiliation{National Astronomical Observatories, Chinese Academy of Sciences, Beijing, 100101, People's Republic of China}
\email{zouhu@nao.cas.cn}

\begin{abstract}\label{abs}
The advancement in sensitivity and field of view of next-generation wide-field survey telescopes requires astrometric measurements with high precision, even in the presence of significant geometric distortions. To address this challenge, we develop a Weighted Polynomial Distortion Correction in 2-Phase (WPDC-2P) method. This approach enhances stellar cross-matching, incorporates distance-based weighting into the traditional polynomial fitting, and employs a look-up table to absorb the remaining distortion residuals. Validated on simulated data from the Main Survey Camera of the \emph{Chinese Space Station Survey Telescope} (CSST), incorporating geometric distortions up to approximately $200$ pixels, the method achieves astrometric standard deviation ranging from 0.013 to 0.107 pixels (0.03 pixels for the $g$-1 detector) across all 18 detectors. Under extreme crowding conditions (e.g., globular cluster NGC 2298), the astrometric precision for the $g$-1 detector reaches 0.05-pixel level within the central region ($r_d < 4000$), despite a centroiding precision of $\sim$0.04 pixels. When applied to the Beijing-Arizona Sky Survey data, for which the standard pipeline delivers an astrometric uncertainty  of $\sim$20 mas, our method reduces the positional scatter to $ \sigma_{\Delta\alpha}=5.494$ mas (0.01 pixels) and $ \sigma_{\Delta\delta}=9.981$ mas (0.02 pixels) using only a weighted 3rd-order polynomial correction. 
The method has been integrated into the CSST data processing pipeline and is prepared for further refinement using on-orbit calibration data.


\end{abstract}


\section{Introduction} \label{sec:intro}

Next-generation survey telescopes are designed to combine higher spatial resolution with wider field of view (FoV), placing increasingly strict requirements on astrometric calibration.
For example, the Legacy Survey of Space and Time \citep[LSST,][]{Ivezic_2019} features a FoV of 3.5$^\circ$ with a pixel scale of 0.20 arcsec, while the space-based Euclid mission \citep{Euclid} offers a visible instrument (VIS) with a FoV of 0.787$^\circ$ and a pixel scale of 0.10 arcsec.
The \emph{Chinese Space Station Survey Telescope}(CSST) \citep{CSST1, Gong_2019, csstcollaboration2025} is a forthcoming space-based survey mission that combines a wide FoV of about 1.1$^\circ$ with high spatial resolution of 0.074 arcsec per pixel. 
It is expected to achieve milliarcsecond-level astrometric performance for billions of celestial sources.

Despite their different designs and observing strategies, geometric distortion (GD) remains a dominant limitation on astrometric accuracy.
Previous studies have demonstrated that high-precision astrometry can be achieved for instruments with relatively small FoV.
For example, the Advanced Camera for Surveys (ACS) Wide Field Channel (WFC) on the Hubble Space Telescope (HST) achieves astrometric precision at the milliarcsecond ($\rm mas$) level \citep{bellini_astrometry_2011} over a 202 arcsec FoV at a pixel scale of 0.05 arcsec. 
Similarly, the Near-Infrared Camera (NIRCam) on James Webb Space Telescope \citep[JWST, ][]{Gardner_2023} provides a 132 arcsec FoV at a pixel scale of 0.03 arcsec, and keeps its astrometric precision of $\sim$0.2 mas for well exposed stellar targets \citep{griggio_photometry_2023}. 
In contrast, high-precision astrometry becomes more challenging for wider-field surveys.
In case of Euclid, the detector is divided into quadrants, each with a FoV of 200 arcsec at the pixel scale of 0.1 arcsec. 
By solving for GD independently within each quadrant, this approach effectively avoids cross-matching issues in the wide FoV. As a result, each quadrant of the VIS on Euclid achieves an astrometric accuracy better than 8 mas \citep{collaboration_euclid_2025}.
The Beijing–Arizona Sky Survey \citep[BASS, ][]{BASS_zouhu}, with a FoV of 1.12$^\circ$ and a pixel scale of 0.45 arcsec, faces challenges due to relatively low charge transfer efficiency (CTE) and differential chromatic refraction (DCR), leading to centroid offset of about 0.1 pixels \citep{Peng_2023}.

GD is an inherent image "warping" caused by imperfections in telescope optics. This distortion alters the true spatial relationships between objects, affecting their visual morphology and relative separations across the field.
Conventional methods for correcting GD are primarily categorized in two ways by their reference frame origin: the self-calibration method, which iteratively refines the reference positions using astrometric residuals with respect to a reference frame \citep[e.g.,][]{HSTWFC2_GD, HSTWFC3_GD, jwst_2024}, and the external calibration method, which directly computes GD parameters from an external catalog with minimal observational overhead \citep{HST_wfc3_GD0, wfc3_GD0}. 
Despite their different approaches, both methodologies rely predominantly on polynomial models to describe the GD, with variations arising in the polynomial format and order. 
For instance, cameras on space telescopes (e.g., HST/ACS, JWST/NIRCam, and Euclid/VIS) consistently employ the 3rd-order TAN-SIP polynomial for its intuitive, pixel-based representation of distortion \citep{bellini_astrometry_2011, griggio_photometry_2023, collaboration_euclid_2025}. 
In contrast, ground-based systems exhibit greater diversity, as seen in  BASS (4th-order TAN-PV) \citep{Peng_2023},  Keck I/OSIRIS (5th-order Legendre) \citep{Keck_1}, Subaru and LSST (9th-order TAN-SIP) \citep{bosch_hyper_2018, Ivezic_2019}.

However, increasing the polynomial order requires a denser and more uniform distribution of reference stars across the field which is usually hard to satisfy in external-calibration case. 
For example, \textsc{SCAMP} \citep{Bertin_2006} and other existing astrometric calibration tools may not perform adequately in the near-ultraviolet band or in sparse stellar fields. 
At the same time, a nonuniform distribution of cross-matched reference stars can lead to localized anomalies in the correction. 
Finally, cross-matching over a wide FoV also imposes a new challenge for external calibration methods, as any errors in this process can directly limit the accuracy of the GD correction.
For self-calibration, more sophisticated algorithms and several appropriately dithered observations are required to mitigate distortion errors in the initial self-constructed reference frame.
Even a sudden minor telescope vibration can cause GD pattern to drift over time, potentially increasing the matching error.
The CSST will observe 17,500 square degrees, covering regions with varying stellar densities, with each exposure time in 150 seconds. 
Its optical design and on-orbit operating conditions produce unique distortion patterns that require a dedicated correction strategy. 
Achieving robust astrometric calibration for such an instrument is essential for the scientific goals of the mission and requires methods that remain stable across the full detector area.

Motivated by these challenges, this work aims to develop and validate a robust GD correction method for the CSST. 
We propose a template-based analysis method, the Weighted Polynomial Distortion Correction in 2-Phase(WPDC-2P).
Section \ref{sec:Data} presents the characteristics of the CSST simulated dataset used in this work and details the construction of the GD model. 
The methodology of WPDC-2P is described in Section \ref{sec:method}, and  its application is demonstrated in Section \ref{sec:application} on two distinct scenarios: a simulated globular cluster with significant detector undersampling and star overlapping, and real observational data from the BASS.
Finally, the limitations of our algorithm  and our conclusions are summarized in Section \ref{sec:discussion and clusion}.

\section{Data} \label{sec:Data}
The CSST is one of the most critical scientific facilities in the China Manned Space Program. 
With a primary mirror aperture of 2 meters, it carries multiple scientific instruments, among which the Main Survey Camera (MSC) is the primary instrument for imaging observations.
The MSC includes seven photometric filters (\textit{NUV, u, g, r, i, z, y}) installed on 18 detectors, along with three spectroscopic gratings (\textit{GU, GV, GI}) on 12 detectors. 
Each detector contains 9216 $\times$ 9232 pixels with a pixel scale of 0.074 arcsec per pixel, corresponding to a FoV of approximately 660 arcsec on a side. 
All 30 detectors together provide a total FoV of about 1.1$^\circ \times 1.1^\circ$ on the sky, making accurate modeling and correction of GD across the full focal plane a critical requirement for astrometric applications.

The GD correction algorithm was developed based on CSST simulated data, which consist of both images and a mock catalog.
The CSST end-to-end simulation software \citep{weichengliang2025} was used to generate simulated observational data covering a 25-square-degree sky area with a variety of pointing locations and stellar number densities.
The simulations incorporate: 
(1) astrophysical source models including deep-field stellar catalogs \citep{wei2025mockobservationscsstmission} , semi-analytic galaxies, gravitational lenses, quasars, and solar system objects; 
(2) full telescope engineering constraints such as mirror fabrication tolerances, mechanical jitter, thermal deformation effects, and ground-to-orbit gravity release residuals, which collectively estimate on-orbit wavefront deviations from the ideal optical design; and 
(3) detector-level realism through convolution of targets with laboratory-measured point spread function (PSF) fields \citep{Banzhang2025}, generating multi-band photon distributions across the focal plane. 
Furthermore, to simulate different observational epochs, it accounts for stellar proper motions and the varying distributions of galaxies and other celestial objects, even within the same sky area.

\begin{figure}[ht!]
\centering 
\includegraphics[width=0.7\linewidth]{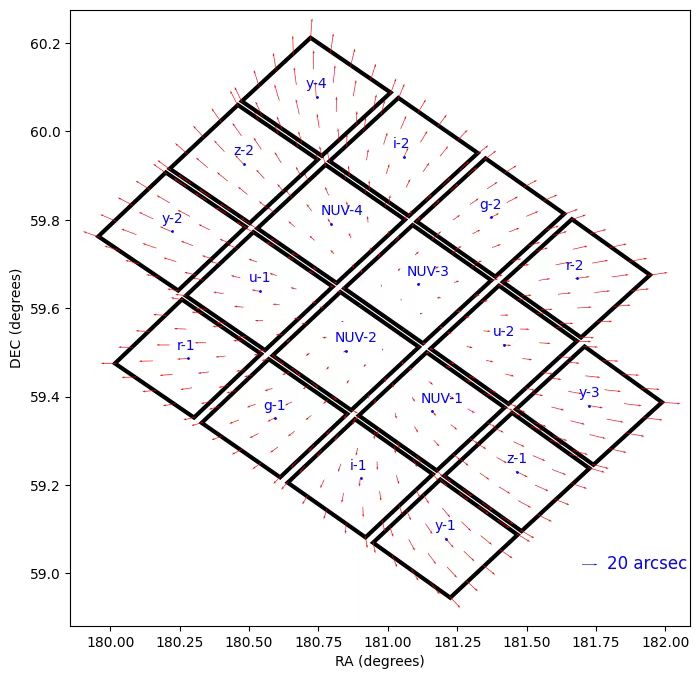} 
\caption{The distortion map of the MSC's 18 photometric detectors, derived from a subset of CSST 25-square-degree simulation program. 
Detector name and number are shown in blue on the central region of each detector box. The size of distortion vector marked in red is magnified by a factor of 10.
}
\label{fig:distortion_map} 
\end{figure}

The simulated GD model was constructed based on  computed and wavelength-dependent PSFs \citep{Banzhang2025}, and subsequently applied to the simulated images. 
The initial World Coordinate System (WCS) parameters, including the coordinates of the reference point and the scale matrix, are stored in the FITS header. For CSST, the central point of each detector is chosen as the reference point, with its celestial coordinates obtained through interpolation. 
Figure. \ref{fig:distortion_map} illustrates the constructed GD model embedded in the simulated data, where the red arrows denote the distortion vectors (magnified by a factor of 10 for clarity).  
The system exhibits a pronounced asymmetric radial distortion, characterized by a positional displacement that increases from the center toward the edges.  
This effect is particularly severe near the corners, where the pixel offset in the $y$-band detectors can reach $\sim$200 pixels (corresponding to $\sim$15~arcsec). 
This embedded GD prior model needs to be derived during the astrometric calibration process.

In addition, the simulated images are used for astrometric measurements, while the mock catalog serves as the reference for evaluating the GD solution.
The mock catalog provides undistorted sky coordinates together with their corresponding distorted detector positions given by the simulated GD model (e.g., pixel coordinates $(X_{\rm ref}, Y_{\rm ref})$ and celestial coordinates $(\alpha_{\rm ref}, \delta_{\rm ref})$), photometric magnitudes, and intrinsic stellar properties. 
Thus this mock catalog represents the ground truth including all deterministic optical and geometric effects and subsequently used as the reference catalog in the cross-matching procedure described in Section~\ref{subsec:star-match}.

\section{GD CORRECTION: WPDC-2P} \label{sec:method}

The GD solution for each optical camera detector of the CSST (comprising four $NUV$ and $y$ detectors, and two detectors for each of the $u$, $g$, $r$, $i$, and $z$ bands) was independently derived by using a two-step procedure.
First, a distance-weighted, 3rd-order polynomial was iteratively fitted to the cross-matched sources from the reference catalog to determine the polynomial variation ($ PV $) coefficients.
Second, a look-up table (LUT) was applied to account for high-frequency spatial systematic residuals that cannot be absorbed by the polynomial alone. 
The GD solution, which includes both the distance-weighted $ PV $ fitting and the LUT, is stored in the FITS header to enable positional calibration during the data processing. 
We refer to this entire procedure as the WPDC-2P method.

The 25-square-degree data set spans both high- and low-latitude regions, with stellar number densities ranging from a few dozen to several thousand per square arcminute. 
We adopt the $g$-1 band detector as a representative example because of its stable image quality and low measurement uncertainties, which are well suited for constructing a high-precision GD model. 
In this work, we use 25 $g$-1 band images from the data set to illustrate our method.

Throughout this paper, positional residuals are defined as $\Delta X = X' - X_{\rm  ref}$ and $\Delta Y = Y' - Y_{\rm ref}$ (see details in Section \ref{subsec:poly-fit}) in pixel coordinates. 
Astrometric precision here is quantified by the $1\sigma$ standard deviation of the residual distribution derived from ensembles of multiple images of a same detector. 
Astrometric accuracy is evaluated with respect to different reference catalogs in different sections: the mock catalog is used as the reference  for simulation data in sections \ref{sec:method} and \ref{subsec:NGC 2298}, while an external reference catalog (Gaia DR3) is adopted in section \ref{subsec:bass}.
In the simulated experiments, the mock catalog is treated as the noise-free ground truth; therefore does not include uncertainties in coordinates, proper motions, or parallaxes.
As a result, the derived astrometric accuracy represents the intrinsic performance limit of the GD correction method under ideal reference conditions.

\subsection{Stellar Cross-Matching} \label{subsec:star-match}
Stellar positions and fluxes in each CSST simulated images were measured through python $PhotUtils$ software \citep{photutils_larry}.
Detector undersampling and $\mathrm{S/N}$ are major challenges for achieving high-precision photometry and astrometry.
Based on our tests using these undersampled CSST simulated point-source images, we find that PSF-fitting delivers more accurate centroid estimates than simpler approaches such as the moment method or windowed centroiding.

\begin{figure}[ht!]
\centering 
\includegraphics[width=0.95\linewidth]{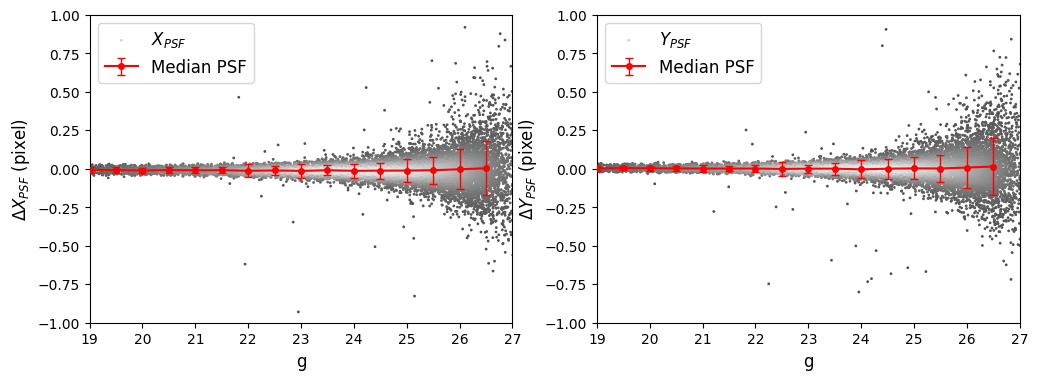} 
\caption{ Centroiding residuals as a function of source magnitude for 25 simulated CSST $g$-1 images with an exposure time of 150 seconds. The left and right panels show $\Delta X_{\rm PSF} = X_{\rm PSF} - X_{\rm ref}$ and $\Delta Y_{\rm PSF} = Y_{\rm PSF} - Y_{\rm ref}$, respectively, where $X_{\rm PSF}$ and $Y_{\rm PSF}$ are the centroid positions measured from PSF fitting, and $X_{\rm ref}$ and $Y_{\rm ref}$ are the corresponding simulated truth positions from mock reference catalog. Gray points represent individual stars, while red solid lines with error bars indicate the mean offsets and the corresponding $1\sigma$ centroiding precision in 0.5 mag bins.
}
\label{fig:xy_off_obs} 
\end{figure}

Fig \ref{fig:xy_off_obs} shows the dependence of centroiding residuals on source magnitude under simulated undersampling conditions for the MSC, assuming an exposure time of 150 seconds per image.
For bright stars ($g < 20$), the centroiding precision is approximately 0.02 pixels in both X and Y. 
As the magnitude increases to 25, the precision gradually degrades to 0.07 pixels.
For fainter sources $g > 25$, the dispersion of the residuals exceeds 0.1 pixels, surpassing the level required for reliable GD calibration.
Therefore, to ensure the quality of the correction, only stars $g \leq 24.5$ are considered.

\begin{figure}[ht!]
\centering 
\includegraphics[width=0.95\linewidth]{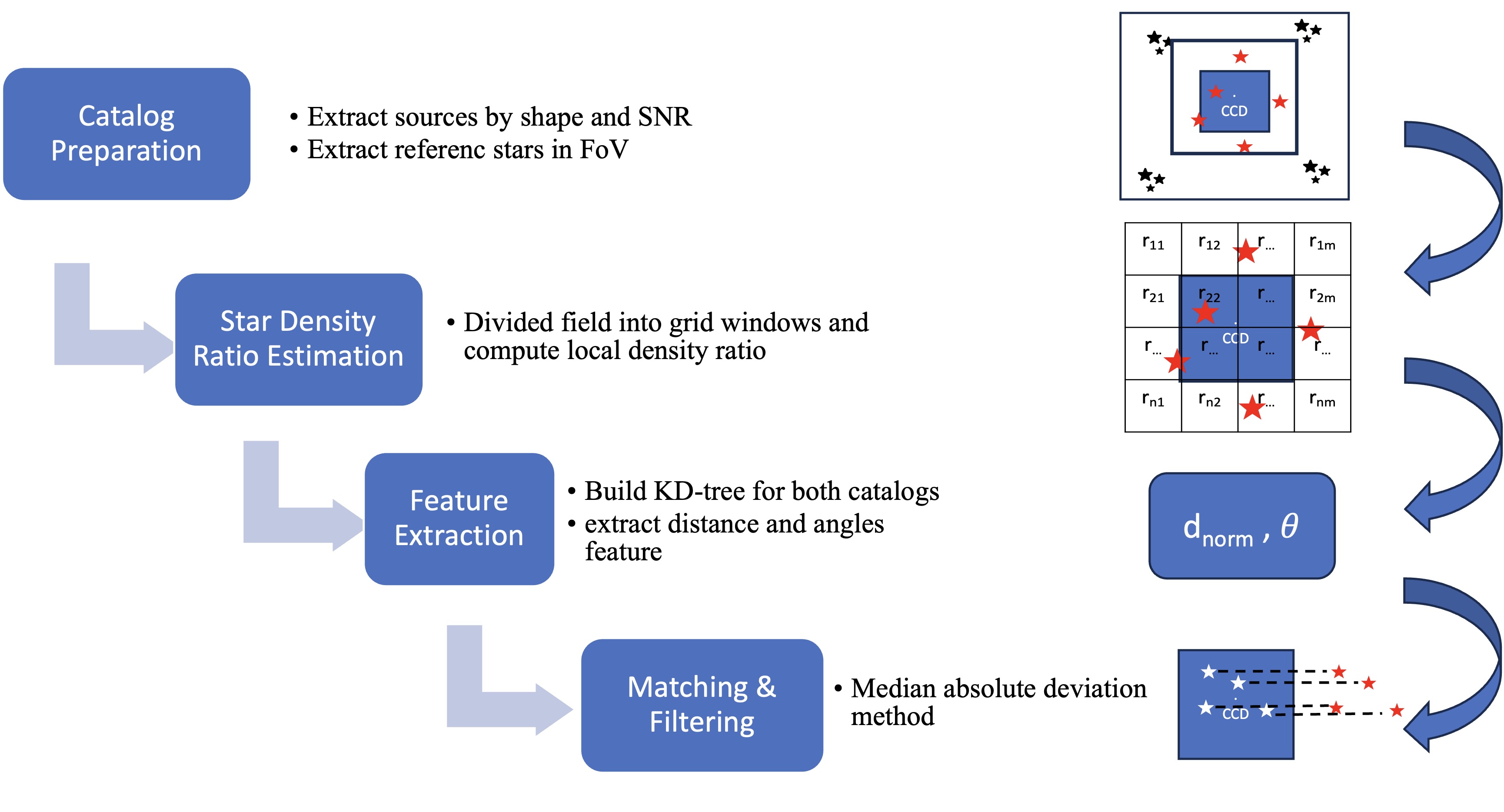}   \hspace{-0.5cm}
\caption{Workflow of the DiGStar algorithm for cross-matching. The process involves: (1) Data clipping and source extraction, (2) Grid-based local star density ratio analysis of the image, (3) KD-tree construction for efficient neighbor searches and relative position feature extraction, (4) Two-stage matching with outlier rejection (MAD filtering). Finally merging matched pairs with source metadata.} \label{fig:match_work_flow} 
\end{figure}

We employ DiGStar \citep[in preparation]{Wu2025DiGStar}, a density-guided, geometrically invariant, multi-stage stellar cross-matching algorithm.
By constructing local relative-position features and executing a coarse-to-fine matching pipeline, it maintains high accuracy and computational efficiency even in crowded stellar fields. 
The overall workflow, as depicted in Figure. \ref{fig:match_work_flow}, encompasses catalog preparation, stellar density ratio estimation, feature construction, multi-stage source matching, and outlier rejection. 
Together, these steps produce a consistently cross-identified catalog of matched sources.

The detailed procedure is described below.
\begin{itemize}
\item[1).] \textit{Catalog Preparation}
Reference stars were selected from the reference catalog within an approximately $20$ arcmin FoV centered on the image reference point.
Measured image stars were further filtered to improve matching efficiency by applying the following criteria:
\begin{itemize}
    \item Rejecting extended sources with $|roundness|>$0.3, as they are likely galaxies.
    \item Requiring $\mathrm{S/N}$$>$10 to ensure reliability. 
\end{itemize}
    
\item[2).] \textit{Star Density Ratio Estimation} We estimate the local ratio between the measured stellar-candidate catalog and the reference catalog by dividing the field into a grid and computing the density ratio within each cell. 
This map of local ratios guides the adaptive selection of neighborhood sizes in subsequent matching stages, improving matching accuracy and robustness in dense or  spatially non-uniform fields.  
\item [3).] \textit{Feature Extraction}
Once the density ratio map is obtained, a KD-tree structures \citep{KDtree_Bentley} for both measured and reference catalogs are constructed. 
These structures are used to determine the  relative positional features like normalized distances $d_{\mathrm{norm}}$ and angular orientations $\theta$.
For each star in both catalogs, a set of relative position features is generated based on its local stellar neighborhood, and then stored as ID-indexed arrays to support rapid similarity comparisons.

\item[4).] \textit{Matching \& Filtering} The matching process is initially carried out using these relative position features ($d_{\mathrm{norm}},\theta$) via KD-tree queries, with a distance threshold that is adaptively adjusted to ensure sufficient candidate pairs. 
To minimize false positives, initial matching result is filtered using the Median Absolute Deviation (MAD) method, followed by thresholding based on positional residuals. 
A subsequent fine matching step, with a slightly relaxed threshold, captures potentially missed targets.
\end{itemize}

After the cross-matching procedure, we obtain a matched catalog that contains the celestial coordinates $(\alpha_{\rm ref}, \delta_{\rm ref})$ and pixel coordinates $(X_{\rm ref}, Y_{\rm ref})$ from the reference catalog , together with the corresponding measured detector coordinates $(X_{\rm PSF}, Y_{\rm PSF})$ for each matched source.
The measured positions $(X_{\rm PSF}, Y_{\rm PSF})$ and matched $(\alpha_{\rm ref}, \delta_{\rm ref})$ are then used as the primary input to solve for the GD parameters. 
The $(X_{\rm ref}, Y_{\rm ref})$ are used latter in section \ref{subsec: LUT} for the LUT calculation.

\subsection{Distance Weighted Polynomial fitting}\label{subsec:poly-fit}

The GD correction employs the $ PV $ distortion \citep{Calabretta} representation with a 3rd-order polynomial model. 
This $ PV $ polynomials provide transformations between distortion-uncorrected intermediate coordinates $(U, V)$ and distortion-corrected intermediate coordinates $(U', V')$. 
The $ PV $ coefficients are stored in FITS header with keywords $\texttt{PVi\_j}$, where $i= 1,2$ are axis indices, and the index $j= 0,1,...,10$ specifies indices of the numeric values for the terms in the polynomials.
Since \textsc{SCAMP} does not compute radial terms defined as $d = \sqrt{U^2 + V^2}$, we exclude them from our model. The $ PV $ polynomials can then be rewritten without explicit dependence on $d$ as follows:

\begin{align}\label{eq_pv3}
    U' = &PV1\_0 + PV1\_1\times U + PV1\_2\times V + PV1\_4\times U^2 + PV1\_5\times U\times V + PV1\_6\times V^2 + \nonumber\\ &PV1\_7\times U^3 + PV1\_8\times U^2 \times V + PV1\_9\times U\times V^2 + PV1\_10\times U^3.\nonumber\\
    \\
    V' = &PV2\_0 + PV2\_1\times V + PV2\_2\times U + PV2\_4\times V^2 + PV2\_5\times U \times V + PV2\_6\times U^2 + \nonumber \\&PV2\_7\times V^3 + PV2\_8\times U\times V^2 + PV2\_9\times U^2\times V + PV2\_10\times V^3.\nonumber
\end{align}

The intermediate coordinates $U$ and $V$ are derived from the relative coordinates ($X, Y$) with respect to the detector's central reference point through the linear transformation:
\begin{gather}\label{eqCD}
\left(
\begin{array}{c}
U\\
V 
\end{array}
\right)= \left ( \begin{array}{cc}
   CD1\_1  & CD1\_2 \\
   CD2\_1  & CD2\_2
\end{array}\right)  
\left(
\begin{array}{cc}
X\\
Y 
\end{array}
\right).
\end{gather}
where the matrix elements $CD~i\_j$ are obtained from the corresponding FITS header keywords, hereafter referred to as the coordinate description matrix ($CD$).
The relative coordinates are defined as $(X,Y) = (X_{\rm PSF}-X_0,Y_{\rm PSF}-Y_0)$, where ($X_{\rm PSF},Y_{\rm PSF}$) are the original detector coordinates measured by PSF-fitting and ($X_0, Y_0$) denote the reference point on each detector, as specified by the FITS header keywords \texttt{CRPIXi}.

Distortion-corrected pixel coordinates $X'$ and $Y'$ are calculated via inverse transform:
\begin{gather}\label{eqCD_inv}
\left(
\begin{array}{c}
X'\\
Y' 
\end{array}
\right)= 
CD^{-1}
\left(
\begin{array}{cc}
U'(\theta_u)\\
V'(\theta_v)
\end{array}
\right).
\end{gather}
where $\theta_u$ and $\theta_v$ denote the sets of polynomial coefficients $\texttt{PVi\_j}$ in Eq. \ref{eq_pv3}.

After the stellar cross-matching procedure is finished, each cross-identification generates a pair of position residuals $\Delta U = U_{\rm ref} - U'$ and $\Delta V = V_{\rm ref} - V'$, where $(U_{\rm ref}, V_{\rm ref})$ are the distortion free intermediate coordinates and can be analytically transformed from the matched reference celestial coordinates ($\alpha_{\rm ref},\delta_{\rm ref}$). 

Considering the difficulty of achieving global correction accuracy with a 3rd-order polynomial function in such a large FoV ($11'\times11'$), we do not adopt higher-order polynomials because the limited number of targets in FoV, especially in $u$ and $NUV$ bands, makes the high-order coefficients poorly constrained and unstable.
Moreover, higher-order GD solutions tend to become unstable and overfit near the field boundaries, and such instabilities would propagate into the LUT and degrade the robustness of the correction.
Therefore, instead of increasing the polynomial order, we first introduce a distance-based weighting function $w(r_d)$, where $r_d$ is the pixel distance from the reference point to a given position in the image.
The optimization objective is to minimize the weighted residuals $\Delta U$ and $\Delta V$ for each object, as expressed in below:

\begin{gather}
\label{eq:objective_func}
\left\{
\begin{array}{c}
{\rm{min}}\sum_i w(r_i)  \cdot \Delta U_i \\
{\rm{min}}\sum_i w(r_i)  \cdot \Delta V_i \\ 
\end{array}
\right.
\end{gather}

The weighting function $w(r_d)$ is defined in a piecewise formulation with continuity enforced at the boundary radius $r_0$, which represents the effective radial extent from the detector center. 
We set $r_0 = 4000$ pixels, approximately half the detector width, while leaving a minimum 500-pixel width at the detector edge to mitigate the influence of edge sources.

Empirical tests indicates that assigning higher weights to the central region ($r_d < r_0$), rather than to edge points, yields more stable polynomial correction in the central field. 
For simplicity, we adopt a uniform weight within the effective radius $r_0$. The weighting function for the inner region is therefore defined as:

\begin{gather}
\label{PR_eq}
    w(r_d\leq r_0) = 1
\end{gather}
For the outer region $(r_d > r_0)$, we evaluate four candidate weighting functions:

\begin{equation}
    \begin{aligned}
        w_1(r_d) &= \exp\left(-\sqrt{r_d-r_0}\right), & w_2(r_d) &=\exp{(-\frac{r_d-r_0}{r_0})^2} .\\
        w_3(r_d) &= \frac{r_0}{r_d}, & w_4(r_d) &= (\frac{r_0}{r_d})^2.
    \end{aligned}
\end{equation}

After applying the weighting functions in Eq. \ref{eq:objective_func}, the $ PV $ coefficients $(\theta_u, \theta_v)$ are obtained via a least-squares fit and used to compute the GD-corrected pixel coordinates $(X',Y')$, with residuals defined as $\Delta X_{\rm PV} = X' - X_{\rm ref}$, and $\Delta Y_{\rm PV} = Y' - Y_{\rm ref}$. 
The cumulative distribution functions (CDFs) $\int_{-1}^{1} p(\Delta X_{\rm PV}) dX$ and $\int_{-1}^{1} p(\Delta Y_{\rm PV}) dY$ are then used to assess the central-field accuracy and to compare weighting functions.
A steep rise near zero indicates most residuals are small, while a gradual slope reflects larger deviations. To reduce the influence of outliers, we quantify the accuracy using $d80$, the width of $\Delta X_{\rm PV}$ (or $\Delta Y_{\rm PV}$) between the 20\% and 80\% CDF levels.

\begin{figure}[htbp!]
\centering 
\includegraphics[width=0.6\linewidth]{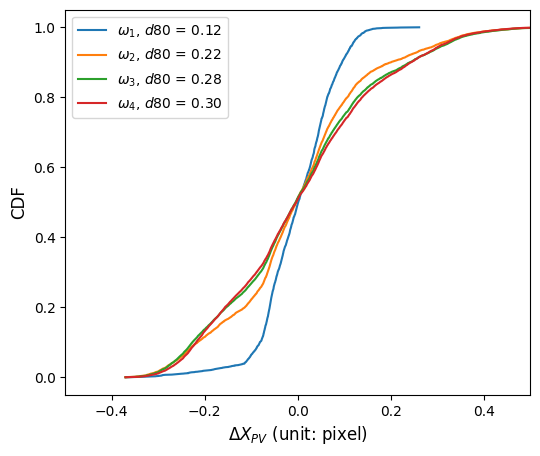} 
\caption{
Cumulative distribution functions of the positional residuals obtained from the four weighting functions in different color.
The accuracy $d80$ denotes the width of $\Delta X$ between the 20\% and 80\% CDF level. The $w_1(r_d)$ scheme shows significantly better performance than the other weighting functions.
}
\label{fig:weight_val} 
\end{figure}

We compute the CDFs of the positional residuals for different weighting functions and compare them in Figure. \ref{fig:weight_val}. 
The CDF for $w_1(r_d)$ rises steeply, with 60\% of residuals within 0.1 pixels, indicating that most positional offsets are very small. Based on this, we adopt $w_1(r_d)$ as the optimal weighting function.

The fitting procedure consists of iterative cycles with two main steps : (1) A least-squares fit to Eq. \ref{eq:objective_func} is performed to update the $ PV $ coefficients, which are incrementally added to those from the previous iteration.
To ensure stable convergence, only 50\% of the newly derived correction is applied to the stellar positions at each step. (2) Using the updated coefficients $(\theta_u, \theta_v)$, the distortion-corrected positions $(X', Y')$ are re-matched to the stars in the reference catalog.
This iterative process is repeated for up to ten iterations or until the positional residuals converge.

\begin{figure}[ht!]
\centering 
\includegraphics[width=0.95\linewidth]{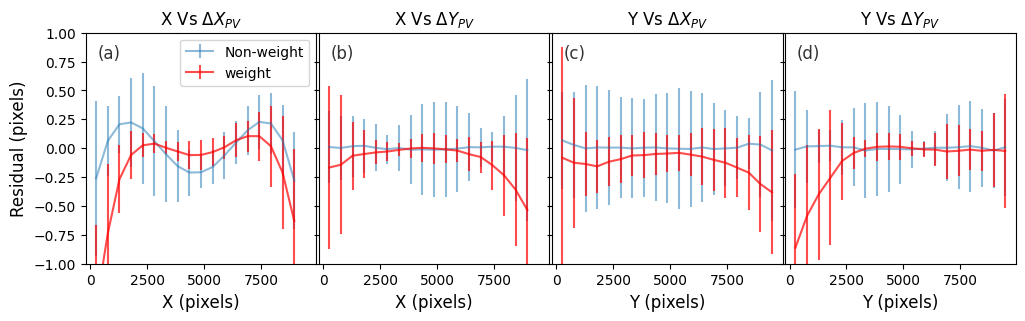} 
\caption{
Positional residuals $(\Delta X_{\rm PV}, \Delta Y_{\rm PV})$ obtained after the 3rd-order $ PV $ parameter fitting. The blue and red points represent the mean residuals of non-weighted and weighted solutions, respectively, computed in slices of 512 pixels. The corresponding blued and red error bars are defined as the 68.27th percentile of the residual distribution (after a $3\sigma$ clipping).
However, overweighting the central region results in systematic deviations of the GD correction at the edges, with larger central-to-edge weight differences producing greater accuracy errors.
All the sources shown have $ \mathrm{S/N} > 10$ in the detector $g$-1 detector.}
\label{fig:weight_effect} 
\end{figure}

A comparison of residual distributions between weighted and unweighted GD solutions is shown in Fig \ref{fig:weight_effect}. 
From the plot of residual distributions, the weighted solution clearly outperforms the unweighted solution. 
In particular, the positional residuals in both the $X$ and $Y$ directions are significantly smaller in the most central area, where the residuals fall below 0.1 pixels. 
However, the limitations of assigning higher weights to central stars are also evident: residuals near the detector edges have a systematic deviations. 
Increasing the polynomial order can reduce these edge residuals to some extent, but this approach requires a substantially larger number of matched stars to effectively constrain the higher-order terms. 
Such a requirement is not feasible in sparse stellar fields, where the number of available reference stars is insufficient. 
Therefore, to address the residual structures that cannot be absorbed by a 3rd-order polynomial, especially in regions with low stellar density, an additional method that does not rely on the availability of numerous reference stars must be pursued.

\subsection{Look-Up Table of Residual Structure}\label{subsec: LUT}

The dispersion of positional residuals $(\Delta X_{\rm PV}, \Delta Y_{\rm PV})$ increases toward the outer regions of the detector, where a 3rd-order polynomial is insufficient to capture the residual distortion. 
To address this limitation, we construct a residual LUT using 25 independent fields.
The combined star counts from these fields in the detector plane yield a high effective source density ($\geq$2000 per square arcmin) with uniform spatial coverage.

The LUT is generated in two steps. First, retrieve positional residuals $(\Delta X_{\rm PV}, \Delta Y_{\rm PV})$ of all the detectors. 
Second, the positional residuals from above are interpolated onto a $50 \times 50$ grid using Python's \texttt{RBFInterpolator} package.
The grid coordinates $(X_{\rm grid}, Y_{\rm grid})$ and their interpolated values $(\Delta X_{\rm LUT},\Delta Y_{\rm LUT})$ together thus form the LUT. 
Each detector has a unique LUT, stored as a FITS binary table extension for future residual correction. An example for the $g$‑1 detector is presented in Table \ref{tab:LUTs}.

{
\scriptsize
\renewcommand{\arraystretch}{0.8}
\begin{table}[htbp]
  \centering
  \caption{Example of $g$-1 detector LUT (Units: Pixels)}
  \label{tab:LUTs}
  \begin{tabular}{llll}
    \toprule
    $X_{\rm grid}$ & $Y_{\rm grid}$ & $\Delta X_{\rm LUT}$ & $\Delta Y_{\rm LUT}$   \\
    \midrule
    0       & 0    & 5.157  & 2.848 \\
    188.082 & 0    & 4.436  & 2.563 \\
    376.163 & 0    & 3.799  & 2.296 \\
    564.245 & 0    & 3.235  & 2.048 \\
    ...     & ...  & ...    & ...    \\
    5642.448     & 4521.796  & 0.000     & 0.055     \\
    5830.531     & 4521.796  & -0.018    & 0.051     \\
    6018.612     & 4521.796  & -0.036    & 0.046     \\
    6206.694     & 4521.796  & -0.053    & 0.039     \\
    ...     & ...  & ...    & ...    \\
    8651.755     & 9232  & 0.405     & -0.899     \\
    8839.837     & 9232  & 0.540     & -0.943     \\
    9027.918     & 9232  & 0.698     & -0.979     \\
    9216    & 9232 & 0.884 & -1.053  \\
    \bottomrule
  \end{tabular}
\end{table}
}

\begin{figure}[htbp]
\centering 
\includegraphics[width=0.95\linewidth]{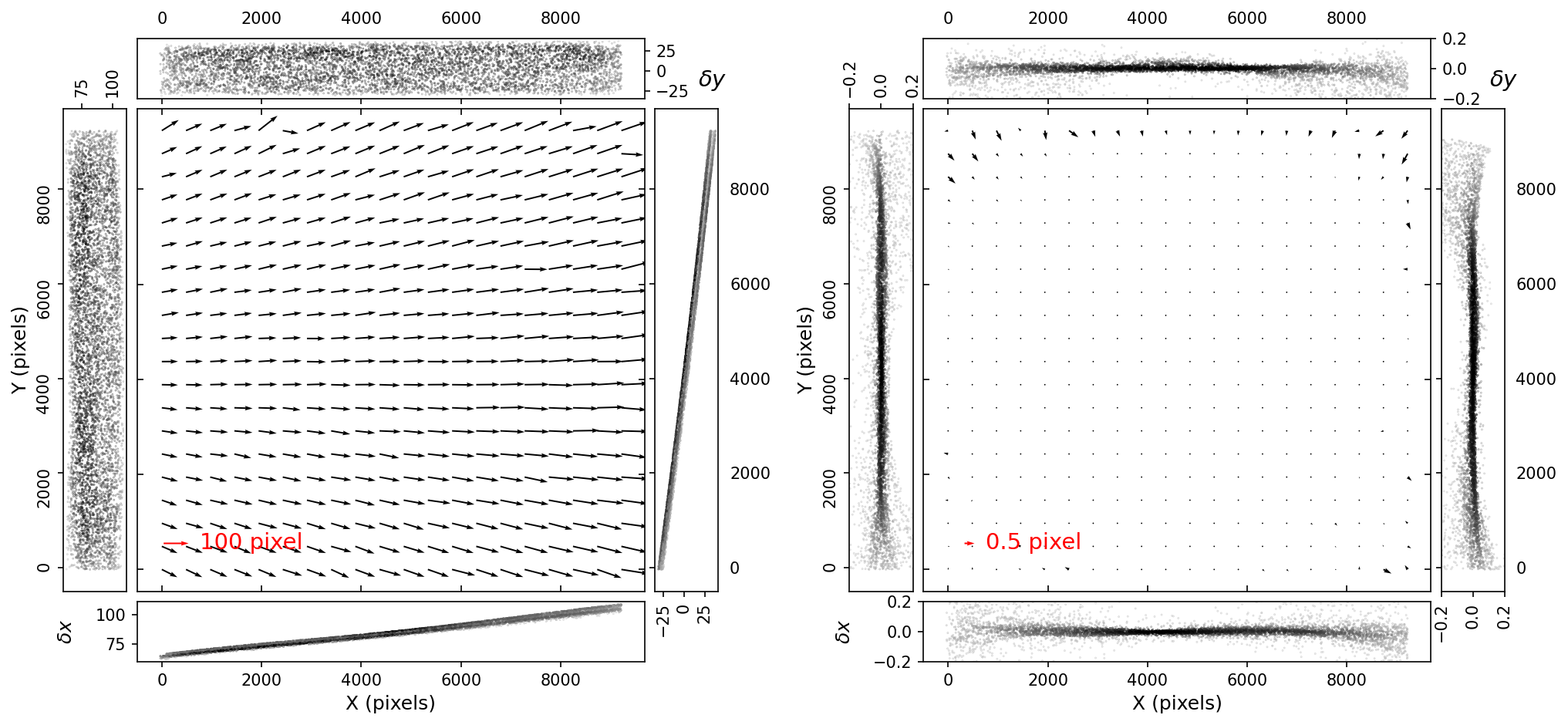} 
\caption{The GD maps for the uncorrected (left) and GD solution accuracy of WPDC-2P result (right). 
All positional offsets are compared with mock reference catalog for 25 different fields observed with the same $g$-1 detector.
In each panel, the two-dimensional residual pattern are averaged over $20\times20$ grid cells and magnified by factors of 5 and 200, respectively.
The surrounding four sub-panels display the $\delta x$ and $\delta y$ residuals as functions of the detector coordinates $(X_{\rm PSF}, Y_{\rm PSF})$, where only 5\% of the residual sample is randomly selected for clarity.
}\label{fig:lut_apply} 
\end{figure}

The application of the LUT relies on the assumption that the GD pattern remains stable. 
Only under this condition can the LUT constructed be applied to other fields. 
Figure. \ref{fig:lut_apply} shows the final astrometric precision, quantified by $\delta x$ and $\delta y$, obtained by applying the distance-weighted 3rd-order polynomial fit combined with the LUT correction to 25 $g$-1 detector images from different fields.

As shown, the final solution achieves an astrometric accuracy better than 0.01 pixels over approximately 90\% of the detector area. 

The corresponding astrometric precision remains well within 0.1 pixels across most of the field, with only a modest degradation toward the detector edges, where the residuals remain below 0.2 pixels.
These results demonstrate that the LUT-based correction strategy is both feasible and reliable. 
Table \ref{tab:csst_lut_params} summarizes the statistics of the final astrometric accuracy and precision for all detectors, including the mean offsets and dispersions after $3\sigma$ clipping.

It is important to note that the GD pattern of a space telescope cannot remain perfectly stable over time as the spacecraft operates in orbit. 
This implies that the GD pattern must be regularly monitored using in-orbit calibration observations. 
Whenever a significant change in the GD pattern is detected, the LUT must be updated using calibration data from the corresponding epoch. 
In addition, GD patterns differ among detectors; for example, both the amplitude and spatial structure of edge-region distortions can vary. As a result, each detector requires its own dedicated LUT, which is not transferable to other detectors.

{
\small 
\renewcommand{\arraystretch}{0.8}
\begin{table}[htbp]
  \centering
  \caption{Positional residual statistics of WPDC-2P for the 18 MSC photometry detectors after $3\sigma$ Clipping (Units: Pixels).  }
  \label{tab:csst_lut_params}
  \begin{tabular}{lllll}
    \toprule
    Detector ID & $\mu_{\delta x}$ & $\sigma_{\delta x}$ & $\mu_{\delta y}$  & $\sigma_{\delta y}$ \\
    \midrule
    y-1  & -0.001  & 0.023 & 0.004 & 0.034 \\
    y-2  & 0.006  & 0.034 & -0.005 & 0.047 \\
    y-3  & -0.006  & 0.047 & -0.005 & 0.051 \\
    y-4  & -0.006  & 0.018 & 0.004 & 0.026 \\
    \midrule
    NUV-1 & 0.005  & 0.013 & 0.003 & 0.021 \\
    NUV-2 & 0.003  & 0.032 & -0.001 & 0.044 \\
    NUV-3 & -0.004 & 0.029 & -0.006 & 0.031 \\
    NUV-4 & -0.010 & 0.034 & 0.007 & 0.085 \\
    \midrule
    u-1 & -0.009  & 0.022 & -0.001 & 0.061 \\
    u-2 & 0.005  & 0.028 & -0.003 & 0.038 \\
    \midrule
    g-1 & -0.004 & 0.029 & -0.005 & 0.025 \\
    g-2 & -0.018 & 0.049 & 0.004 & 0.107 \\
    \midrule
    r-1 & -0.002  & 0.023 & -0.001 & 0.053 \\
    r-2 & -0.008 & 0.055 & 0.016 & 0.052 \\
    \midrule
    i-1 & 0.003  & 0.029 & 0.001 & 0.020 \\
    i-2 & -0.013  & 0.038 & 0.002 & 0.035 \\
    \midrule
    z-1 & 0.002  & 0.018 & 0.002 & 0.026 \\
    z-2 & -0.004 & 0.033 & 0.005 & 0.022 \\
    \bottomrule
  \end{tabular}
\end{table}
}

\section{Application}\label{sec:application}
In this section, we apply the WPDC-2P algorithm to two distinct data sets: a CSST-simulated globular cluster field centered on NGC 2298 and observational data from BASS. 
This serves two purposes: (1) to validate the robustness of the algorithm in crowded stellar fields with significant centroiding errors, and (2) to test its generality across other wide-field surveys.

\subsection{Simulated NGC 2298 Application}\label{subsec:NGC 2298}
\begin{figure}[htbp!]
\centering 
\includegraphics[width=0.95\linewidth]{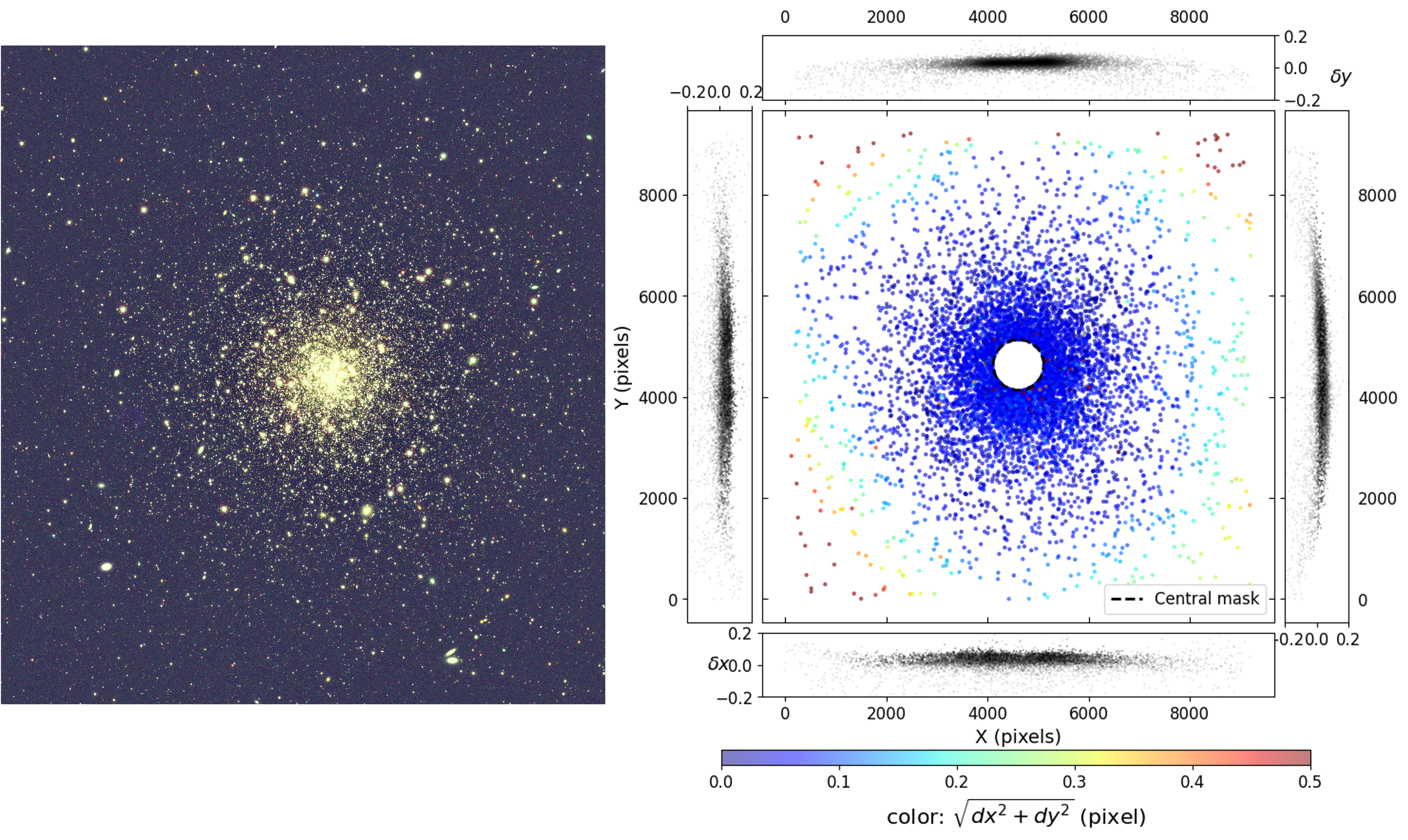} \hspace{0.5cm}
\caption{Left: The color-composite image of the simulated globular cluster NGC 2298. 
Right: Considering the centroiding error, the residual distribution of WPDC-2P solutions for higher $\mathrm{S/N}$ threshold stars ($\mathrm{S/N} > 30$) in $g$-1 band detector. The central white circular region with a radius of 500 pixels is masked due to crowding caused by the extreme stellar density in the cluster core, which leads to centroiding errors exceeding 0.1 pixels. }\label{fig:NGC_2298_test} 
\end{figure}

We select the simulated globular cluster NGC2298 (J2000: $\alpha = 102.246^\circ$, $\delta = -36.005^\circ$) as our test field because it spans a wide range of stellar densities, from a few dozen to over 5000 stars per square arcminute, enabling validation in both sparse and crowded regions. 
We employ the CSST end-to-end simulation pipeline \citep{weichengliang2025} to generate simulated images in seven photometric bands (shown in the left panel of Figure \ref{fig:NGC_2298_test}). 
To ensure the cluster center remained within the central region of all 18 detectors, we used 18 exposures, each with a small pointing offset. 

\begin{figure}[ht!]
\centering 
\includegraphics[width=0.95\linewidth]{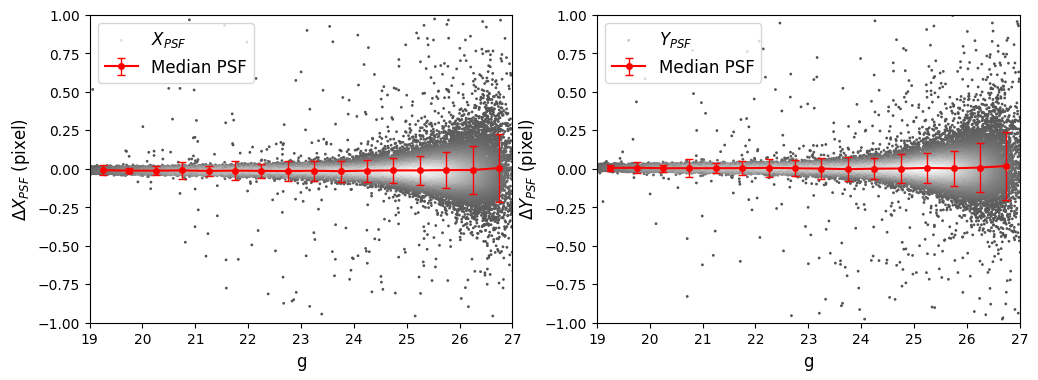} 
\caption{ Same as Figure. \ref{fig:xy_off_obs} but the result of centroiding error for CSST-simulated NGC 2298. The precision of PSF-fitting centroiding had increase a factor of 2 compared with result in 25-square-degree data set. The errors had reached at 0.06 pixels for bright stars ($g < 22$).
}
\label{fig:xy_off_obs2298}
\end{figure}

Due to detector undersampling and crowding, astrometric errors are significantly increased. As shown in Figure. \ref{fig:xy_off_obs2298}, even well-exposed stars (e.g., $g \leq 22$) exhibit centroiding errors of about 0.06 pixels along both the X and Y axes, compared with $\sim$0.03 pixels for the 25-square-degree data set. We then evaluate the performance of our algorithm under such observational conditions.

We applied our algorithm by matching point-source detections with a higher $\mathrm{S/N}$ threshold ($\mathrm{S/N} > 30$) to the reference catalog.
Taking the $g$-1 detector as an example, the results in the right panel of Figure. \ref{fig:NGC_2298_test} show a slight increase in the dispersion of $(\delta x,\delta y)$ compared with those in Figure. \ref{fig:lut_apply}.
Due to severe crowding in the cluster core $(r_d< 500 \rm pixels)$,  the centroiding dispersion exceeds 0.1 pixels in this region. 
Consequently, the cluster core is masked in the subsequent statistical analysis. 
In the outer region $(500<r_d< 4000)$, the mean positional offsets are $(\mu_{\delta x}, \mu_{\delta y}) = (-0.016, 0.029)$ pixels, with corresponding dispersions of $(\sigma_{\delta x}, \sigma_{\delta y}) = (0.051, 0.053)$ pixels. These values are slightly larger than those obtained from the 25-square-degree data set for the $g$-1 detector.
These results indicate that the additional random centroiding errors introduced by crowding can be effectively absorbed by either the distance-weighted polynomial model or the LUT correction, even in a dense stellar field. 
However, they also demonstrate that the final astrometric accuracy is ultimately limited by the intrinsic precision of the centroiding process itself. For example, \citet{Nie_Ap} developed a multi-Gaussian PSF-fitting algorithm that achieved astrometric accuracies of $\sim$1 mas for stars with $g \leq 20$ under sparse stellar field in the context of CSST.

\begin{figure}[htbp!]
\centering 
\includegraphics[width=0.95\linewidth]{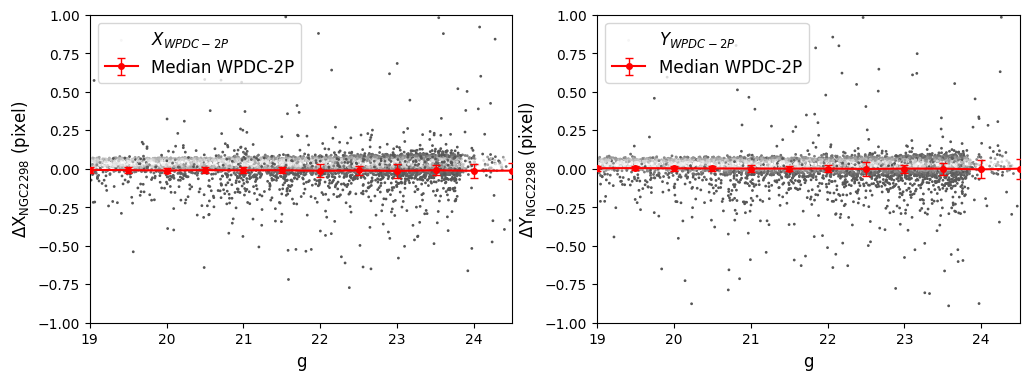}  \hspace{-0.5cm}
\caption{Distribution of GD-corrected positional residuals $(\Delta X_{\rm NGC2298}, \Delta Y_{\rm NGC2298})$ as a function of $g$-band magnitude. The grayscale shading represents the local density of points, with darker shades indicating higher densities. The red solid lines with error bars show the mean positional offsets and the corresponding $1\sigma$ dispersions, computed within 0.5 mag bins.}\label{fig:mag_dxy} 
\end{figure}

In addition to examining the spatial distribution of residuals across the detector, we further investigate the dependence of the solution performance on source magnitude. 
Figure. \ref{fig:mag_dxy} shows the corrected positional residuals $(\Delta X_{\rm NGC2298}, \Delta Y_{\rm NGC2298})$ as a function of magnitude.
The error-magnitude relation indicates that the astrometric precision of the GD solution remains relatively uniform over the magnitude range $19 \leq g \leq 24$, varying from 0.048 pixels at the bright end ($g= 19$) to 0.089 pixels at the faint end ($g = 23.75$).
Compared to the results from the 25‑square‑degree field for $g$-1 detector, this uniform but degraded precision is primarily attributable to the additional centroiding errors introduced in the crowded stellar field.
Notably, the precision does not exhibit the expected significant monotonic decline with decreasing signal-to-noise ratio ($\mathrm{S/N}$). This behavior arises because the fitting process is dominated by the more numerous fainter stars, whose centroiding uncertainties set a nearly uniform error floor across the magnitude range.

\subsection{Beijing–Arizona Sky Survey Application}
\label{subsec:bass}
To validate the general applicability of our algorithm, we further tested it on BASS, which is a wide-field imaging survey covering 5400 square degrees in the northern Galactic cap. 
Conducted with the Bok 2.3 m Telescope at Kitt Peak, BASS provides deep imaging in the $g$ and $r$ bands with a pixel scale of 0.454 arcsec. 
Its astrometric error dispersion varies with celestial coordinates, with a random error of about 16 mas in low-declination regions and up to about 22 mas in the high-declination regions \citep{Peng_2023}. 

For our analysis, we used 10 exposures from the BASS dataset centered at $(\alpha,\delta) = (278.8^\circ, 40.0^\circ)$, taken on 30 April 2017, each consisting of two $g$-band and two $r$-band images.
Point sources were detected and filtered using $PhotUtils$, and then cross-matched with the Gaia DR3 reference catalog using the matching algorithm described in Section \ref{subsec:star-match}. 
A total of 49,158 sources were obtained for statistical analysis.
The $ PV $ coefficients were subsequently derived through an iterative fitting procedure described in Section  \ref{subsec:poly-fit}. 
Given that each BASS detector has a size of 4k pixels, we set the distance threshold to $r_0 = 2000 ~\rm pixels$. 
Unlike the previous test on NGC 2298, where the reference point lies within the detector area, the Bok telescope’s reference points are located outside the detectors (at positions symmetric about the four detector centers). 
Therefore, for each detector, we selected the matched source nearest to the detector center as the reference point to define a local reference coordinate frame, which was then used for subsequent GD correction.

Because the these observations are taken only about one year from the Gaia DR3 reference epoch (2016), Gaia positional and proper-motion propagation errors are typically at the sub-mas to 1 mas level for $g \lesssim 21$\citep{gaia_edr3}, implying that the reported astrometric errors are dominated by the BASS measurements and the GD calibration, rather than by uncertainties in the Gaia reference catalog.

\begin{figure}
\centering 
\includegraphics[width=0.95\linewidth]{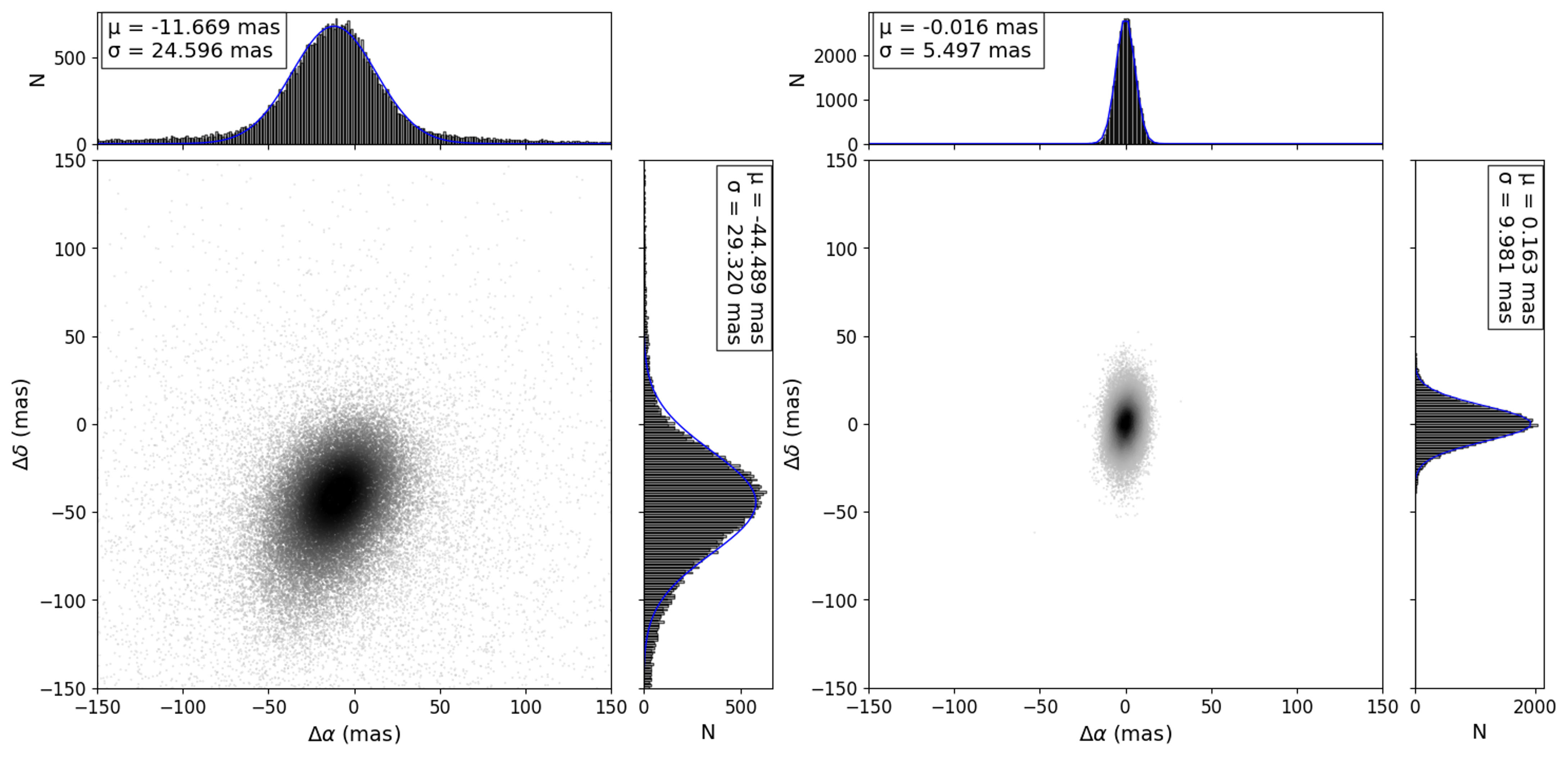} \hspace{-0.5cm}
\caption{ Performance comparison of GD correction results on 10 exposures (each with two $g$-band images and $r$-band images) of the BASS dataset with Gaia DR3.
Left: GD solution of traditional \textsc{SCAMP} result in released catalogs with 48,256 matched stars. Right: Our method with 49,158 matched stars.
The weighted polynomial corrected stellar positions achieving a mean error $( \mu_{\Delta\alpha}, \mu_{\Delta\delta}) = (0.016,0.163)$ mas and standard deviation $(\sigma_{\Delta\alpha}, \sigma_{\Delta\delta}) = (5.494, 9.981)$ mas, compared to the original BASS pipeline results $( \mu_{\Delta\alpha}, \mu_{\Delta\delta}) = (-11.669, -44.489)\rm ~mas,(\sigma_{\Delta\alpha}, \sigma_{\Delta\delta}) = (24.596, 29.320)$ mas. 
The side and top panels show the histograms of $\Delta\alpha$ and $\Delta\delta$, respectively, with Gaussian fits overlaid.}\label{fig:Bass_result}
\end{figure}

Furthermore, because of the large number of charge transfers along the readout direction, the CTE significantly varies with both position and signal level along $y$-axis. 
Compared with the traditional \textsc{SCAMP} processed results, our method achieves mean astrometric offsets of -0.016 mas and 0.163 mas in $\alpha$ and $\delta$ directions, respectively.
The corresponding astrometric precision, quantified by $(\sigma_{\Delta\alpha}, \sigma_{\Delta\delta})$, is reduced from $(24.596, 29.320)$ mas to $(5.494, 9.981)$, representing an improvement in precision by a factor of $\sim$3 (see in Figure.~\ref{fig:Bass_result}). 
This means that a weighted 3rd-order polynomial correction alone achieves a residual precision better than 0.02 pixels.
For comparison, interferometric calibration techniques can determine detector pixel positions to a precision down to $\sim 0.01$ pixel \citep{crouzier_2016, Shao_2023}, indicating that our results are approaching the instrumental limit.
These results demonstrate the robustness of our method under real-world wide-field conditions, even in the absence of prior instrument-specific distortion models.

\section{Conclusion}\label{sec:discussion and clusion}

In this paper, we have investigated a GD correction method, WPDC-2P, a GD correction method designed for high-resolution wide-field astrometric surveys. 
The algorithm is characterized by the following key features:
\begin{enumerate}
\item It employs position features $(d_{norm}, \theta)$, which describe the normalized distances and relative angular orientations between each star and its neighboring stars, to perform cross-matching with reference catalogs, thereby significantly increasing the number of usable reference stars.

\item A distance-based weighting scheme is incorporated into the traditional polynomial fitting, enabling a 3rd-order model to achieve $\sigma_{\Delta X}, \sigma_{\Delta Y} < 0.1$ pixels in the central region of the detector, with only a slight loss of precision toward the edges.

\item The astrometric precision across the full field of view is homogenized and improved through the use of a LUT, which absorbs residual distortions near the detector edges and further refines the solution in the central region. 
The combination of $ PV $ fitting and a dynamic LUT establishes a continuously refined internal astrometric reference frame.
This self-correcting capability greatly reduces the need for calibration-field observations.

\end{enumerate}

The method is applied to CSST MSC simulated data, which comprise 18 detectors with a pixel scale of 0.074 arcsec per pixel and collectively cover a FoV of approximately 1 square degree. 
With WPDC-2P, the solution achieves mean positional offsets  $\mu_{\Delta X},\mu_{\Delta Y} <0.02$ pixels and dispersion $\sigma_{\Delta X},\sigma_{\Delta Y}$  ranging from 0.013 to 0.107 pixels across all detectors. 
Its robustness was further validated in a simulated crowded field (NGC 2298) incorporating centroiding errors, where the error dispersion reaches $\sigma_{\Delta X} = 0.051$ pixels and $\sigma_{\Delta Y} = 0.053$ pixels in the central cluster region ($r_d < 4000$).
The algorithm also demonstrated general applicability to data from the BASS survey.
Using the distance-weighted polynomial correction alone, dispersions of 5.494 mas (0.01 pixels) in $\alpha$ and 9.981 mas (0.02 pixels) in $\delta$ are achieved after cross-matching with Gaia DR3.

The present method has several limitations.
Foremost among these are undersampling and low $\mathrm{S/N}$, which directly increase centroiding errors and  thus limit GD correction precision. 
This limitation establishes an effective error floor: for stars brighter than 25th magnitude (e.g., $g<25$), the dispersion of PSF-fitting centroiding errors reaches 0.06 pixels, comparable to the accuracy of LUT corrections. Consequently, the number of stars that satisfy the precision requirements for a reliable distortion solution is substantially reduced.

The algorithm has been integrated into the CSST simulated data processing pipeline. 
However, the current validation is limited by the use of ground-test parameters, which may not fully reflect on-orbit conditions, such as pointing stability and detector effects. 
In future applications, the algorithm can refine GD parameters and update the LUT on a per-exposure basis, circumventing the need for dedicated calibration fields required by other surveys.

\section{Acknowledgements}
The mock data in this work is created by the CSST Simulation Team, which is supported by the CSST scientific data processing and analysis system of the China Manned Space Project. 
Jundan Nie acknowledges the support of the National Key R\&D Program of China (grant Nos. 2021YFA1600401 and 2021YFA1600400), the National Natural Science Foundation of China (NSFC) (grant No. 12373019), and the science research grants from the China Manned Space Project (grant No. CMS-CSST-2025-A11).
Yibo Yan acknowledges the support by the China Manned Space Project with grant No. CMS-CSST-2025-A21, the National Key Research and Development Program of China (grants 2025YFF0511000) and the National Natural Science Foundation of China (NSFC; grants 12233008). 

\bibliography{refcite}{}
\bibliographystyle{aasjournal}

\end{document}